# Observation of Significant Photosynthesis in Garden Cress and Cyanobacteria under Simulated Illumination from a K Dwarf Star

**Iva Vilović[1*], Dirk Schulze-Makuch[1-4], and René Heller[5]**

[1]Astrobiology Research Group, Zentrum für Astronomie und Astrophysik, Technische Universität Berlin, 10623 Berlin, Germany
[2]GFZ German Research Center for Geosciences, Section Geomicrobiology; 14473 Potsdam, Germany.
[3]Leibniz-Institute of Freshwater Ecology and Inland Fisheries (IGB), Department of Experimental Limnology; 16775 Stechlin, Germany.
[4]School of the Environment, Washington State University, Pullman; Washington, USA.
[5]Max-Planck-Institut für Sonnensystemforschung, 37077 Göttingen, Germany.
*Corresponding author. Email: iva.vilovic@campus.tu-berlin.de

**Stars with about 45 to 80% the mass of the Sun, so-called K dwarf stars, have previously been proposed as optimal host stars in the search for habitable extrasolar worlds. These stars are abundant, have stable luminosities over billions of years longer than Sun-like stars, and offer favorable space environmental conditions. So far, the theoretical and experimental focus on exoplanet habitability has been on even less massive, though potentially less hospitable red dwarf stars. Here we present the first experimental data on the responses of photosynthetic organisms to a simulated K dwarf spectrum. We find that garden cress *Lepidium sativum* under K-dwarf radiation exhibits comparable growth and photosynthetic efficiency as under solar illumination on Earth. The cyanobacterium *Chroococcidiopsis* sp. CCMEE 029 exhibits significantly higher photosynthetic efficiency and culture growth under K dwarf radiation compared to solar conditions. Our findings of the affirmative responses of these two photosynthetic organisms to K dwarf radiation suggest that exoplanets in the habitable zones around such stars deserve high priority in the search for extrasolar life.**

## Introduction

Light is the fundamental energy source for photosynthesis, enabling the synthesis of organic compounds. Over billions of years, photosynthetic organisms have profoundly transformed our planet into a diverse global ecosystem (Kiang et al., 2007a, 2007b). Therefore, understanding any planet in the context of its stellar environment is an essential step in assessing its habitability. The only life we know of so far has developed around our Sun – a G dwarf star in the Harvard spectral classification system with an effective temperature of 5770 K (Williams, 2013). In addition to the search for Earth-like planets around Sun-like stars, much of the focus in the search of extraterrestrial life (Anglada-Escudé et al., 2016; Gebauer et al., 2021; Gillon et al., 2017, 2016; Rajpurohit et al., 2018) has been placed on the smallest, least massive, and coolest type of star. These red dwarf stars (M dwarfs) possess a range of favorable attributes that are conducive to the detection and potential development of life (Adams et al., 2005). It has even been demonstrated that photosynthetic organisms from Earth can grow and photosynthesize under simulated M dwarf radiation, highlighting the potential viability of modified spectra for biological processes (Battistuzzi et al., 2023a, 2023c; Claudi et al., 2020). That being said, planets orbiting M dwarfs face challenges such as intense tidal forces (Barnes et al., 2009; Heller et al., 2011), significant water loss during the super luminous pre-main sequence phase (Luger and Barnes, 2015), and exposure to substantial levels of extreme UV radiation and high-energy particles. These factors can lead to the photochemical destruction of atmospheric biosignatures and atmospheric erosion (Barnes et al., 2009; Scalo et al., 2007).

In comparison, K dwarf stars are the second most abundant star-types in the Milky Way and their lifetimes vastly exceed the expected main-sequence lifetime of our Sun by billions of years (Arney, 2019; Zakhozhay, 2013). They are also magnetically less active than M dwarfs and thus show less frequent radiation outbursts and coronal mass ejections (Arney, 2019; Zakhozhay, 2013). Planets in the respective habitable zones of K dwarf stars experience up to 1000 times lower UV and X-ray fluxes compared to planets that orbit late M dwarf stars (Richey-Yowell et al., 2019). Most importantly, the habitable zones of the majority of K dwarf stars, when considering their masses and temperatures, do not overlap with the range of distances where a planet in orbit would become tidally locked, preventing dramatic climate variations which could be detrimental for the emergence or sustenance of life (Barnes, 2017; Dole and Henry, 1966; Kasting et al., 1993; Peale, 1977). Furthermore, K dwarf stars have shorter pre-main sequence phases – less than 0.1 Gyr in comparison to 1 Gyr for M dwarfs, and their FUV and X-ray fluxes weaken around six times faster than for M dwarf stars, which potentially allows for early habitability (Arney, 2019; Cuntz and Guinan, 2016). All things combined; K dwarfs can be considered the 'Goldilocks stars' in the search for potentially life-bearing planets. A comparison of all parameters between the Sun–Earth and K dwarf–exoplanetary systems used in this work can be found in **Table 2** in the Methods section.

## Results
### Garden cress (*Lepidium sativum*)

We investigated the viability of simulated K dwarf radiation for the photosynthetic processes of garden cress *Lepidium sativum* and the cyanobacterial strain *Chroococcidiopsis sp. 029*. **Figure 1** shows a top view visual growth comparison of garden cress under solar, K dwarf and dark conditions for selected days after sowing the seeds. The garden cress under K dwarf radiation was visually comparable to garden cress under solar radiation, most of the time even sprouting a day or two earlier.

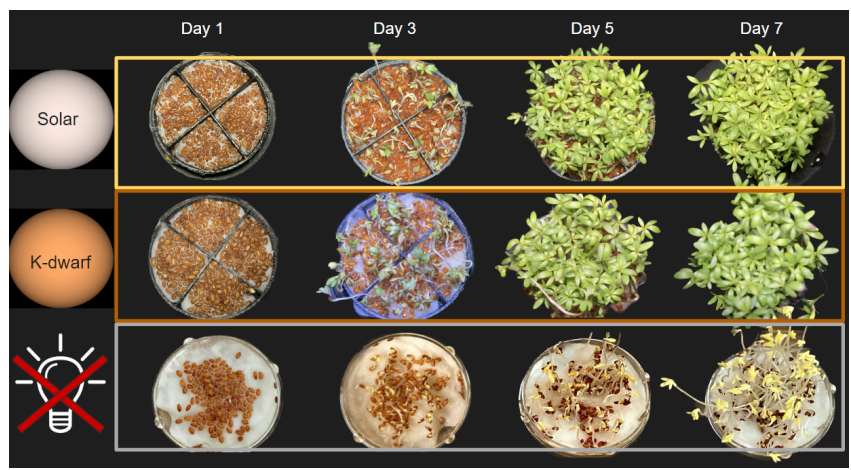

**Figure 1:** Garden cress (*Lepidium sativum*) grown on a sand substrate with one hundred initial seeds under solar (effective temperature 5800 K), K dwarf (effective temperature 4300 K) and dark conditions. Here the visual results for selected days are shown. Garden cress under K dwarf radiation sprouts sooner relative to solar and dark conditions.

It developed a vibrant green color and visual inspection suggests that its leaf surface area appeared to be marginally larger compared to garden cress under solar radiation. As expected, garden cress grown in continuous darkness also sprouted, but it did so with a delay, and it did not develop the green pigment indicative of photosynthesis. Once it sprouted, visual examination suggests that it grew higher than any of the illuminated samples.

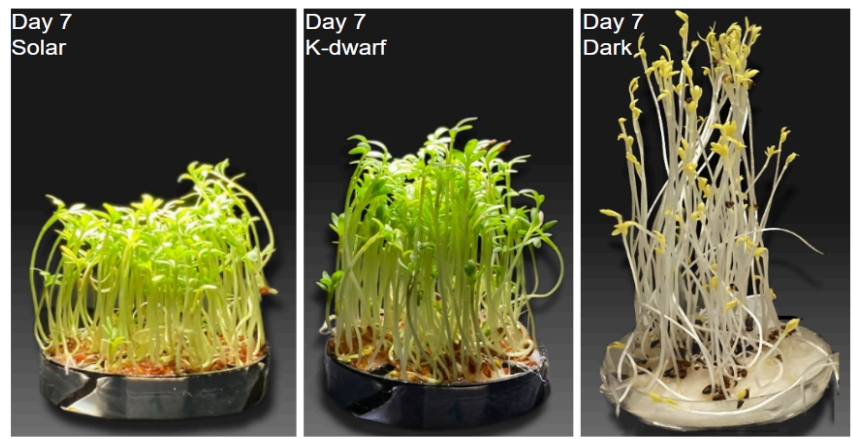

**Figure 2:** Lateral perspective of garden cress (*Lepidium sativum*) grown under solar, K dwarf and dark conditions on the seventh day after sowing.

A lateral perspective of the plant on the seventh day after seed sowing is shown in **Fig. 2.** Visual assessment suggests that there are notable differences in stem elongation and overall height of the garden cress grown under the three irradiation regimes. For example, observations suggest that garden cress under K dwarf radiation grew higher than under solar light. As expected, the plant under constant darkness grew the highest, but the seedlings that did sprout developed a ghostly white/yellow color and grew leaves which folded onto themselves, rather than opening towards the light source like in the irradiated samples.

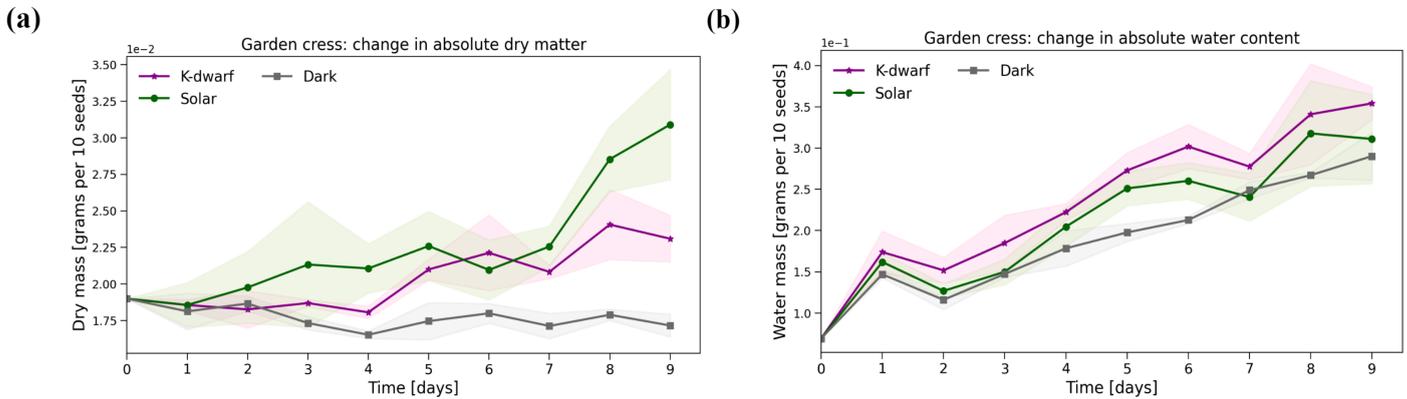

**Figure 3: Dry mass accumulation and water absorption of garden cress under solar, K dwarf and dark conditions with shaded standard deviations. (a) Dry mass accumulation of garden cress measured by desiccating the extracted material. (b) Absolute water contents of garden cress, calculated by subtracting the final dry masses from the initial fresh masses of the extracted material. The error bar is 1σ.**

**Figure 3 (a)** shows the incremental accumulation of dry mass of garden cress under solar, K dwarf, and dark conditions. The dry mass of garden cress has an increasing trend under solar and K dwarf radiation. Compared to this, garden cress under dark conditions fails to accumulate dry mass over time. In fact, its dry mass decreases with respect to initial values. Garden cress under solar conditions exhibits slightly higher, but not significant, absolute dry masses compared to K dwarf conditions ($p > 0.05$, see **Table S1**). The absolute water contents of garden cress under the different radiation regimes are shown in **Fig. 3 (b)**. Garden cress under K dwarf light exhibits marginally higher, but not significant, water contents compared to solar radiation ($p > 0.05$; **Table S1**). Furthermore, garden cress under dark conditions, despite not accumulating any additional dry mass, does consistently absorb water throughout the entire experiment, albeit less than the illuminated plants.

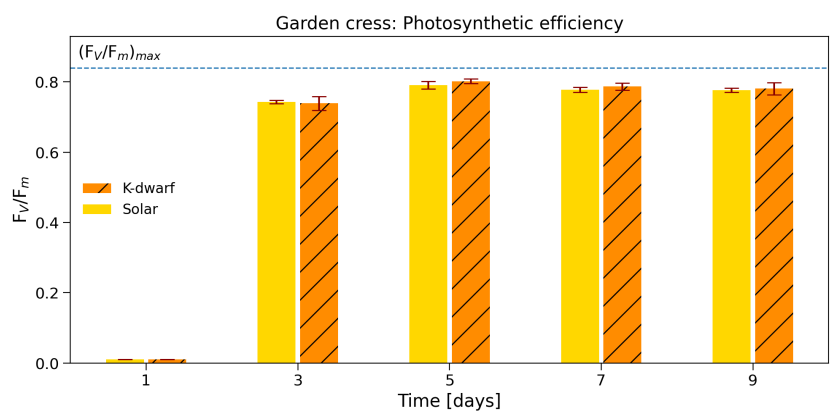

**Figure 4: Photosynthetic efficiency of garden cress under solar and K dwarf conditions on selected days measured with a Pulse-Amplitude-Modulation fluorometer (PAM). $F_m$ is the maximum fluorescence value and is the difference between $F_m$ and $F_0$, where $F_0$ is the minimum fluorescence value. Plant $F_v/F_m$ values range from 0.79 to 0.84, with smaller values indicating plant stress. The error bar is 1σ.**

The photosynthetic efficiencies of garden cress under solar and K dwarf conditions on selected days can be seen in **Fig. 4**. The values for the dark scenario are omitted here, as the registered values were due to photoconversion and not indicative of photosynthesis. The $F_v/F_m$ values for the K dwarf scenario are comparable to those under solar conditions (see **Table 1**). For example, on the last (ninth) day of the experiment, the $F_v/F_m$ values are $0.777 \pm 0.006$ and $0.781 \pm 0.017$ for the solar and K dwarf scenario, respectively. The differences in responses of the illuminated samples are not significant (p > 0.05, see **Table S1**).

**Cyanobacterium (*Chroococcidiopsis* sp. CCMEE 029)**

Due to the growth patterns exhibited by this cyanobacterial strain, the experiment was extended for an additional period of four days compared to the garden cress experiments (Billi et al., 2021). **Figure 5 (a)** shows the phenotypes of the cyanobacterium cultures grown on Agar plates on the final day of the experiment under solar, K dwarf and dark conditions. The cyanobacterium demonstrates a deepening of its green hue under both solar and K dwarf conditions, whereas the cultures under constant dark conditions remained translucent with a light brown edge. **Figure 5 (b)** shows the binary version of the images in **panel (a)**. The pixel values within the regions of interest (ROI) under dark conditions are almost indistinguishable from the background pixels, whereas the pixels of the cultures under solar and K dwarf radiation exhibit a clear edge and filled spots.

The photosynthetic efficiencies of the cyanobacterium from the seventh day of the experiment onward are shown in **Table 1**. Prior to this the culture spots are too diluted for reliable measurements. The photosynthetic efficiencies of the irradiated samples have an overall increasing trend. The maximum fluorescence signal of the K dwarf scenario with the initial Pulse-Amplitude-Modulation fluorometer (PAM) settings was saturated for all but one spot from the tenth day of exposure onwards. This is why only the lower $F_v/F_m$ limit was determined for this scenario (see **Table 1**). Overall, the cyanobacterium exhibited a significantly better response to K dwarf radiation in comparison to solar radiation (p < 0.05, refer to **Table S1**).

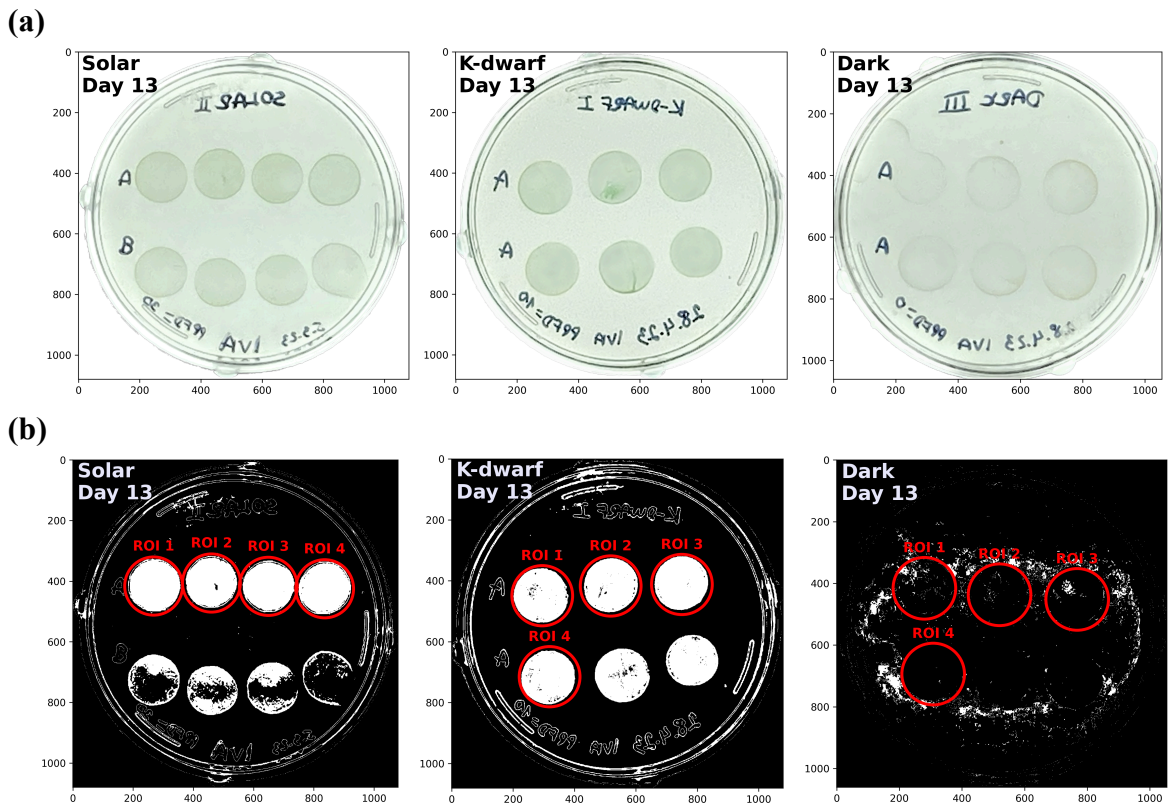

**Figure 5:** Cyanobacterium *Chroococcidiopsis* sp. CCMEE 029 on BG-11 Agar plate under solar, K dwarf and dark conditions. (a) Day 13 after the start of the experiments. The drops correspond to an initial optical density of ~0.7 at 730 nm. The solar Agar plate contains row "B," which corresponds to an initial optical density of ~0.5, and which was not used in the analysis. (b) Binary version of the images in (a). White pixels have been selected through a color threshold. The red circles depict the regions of interest (ROI), i.e., the drops for which the photosynthetic efficiencies and culture densities were determined.

The $F_0$ incremental ratio, i.e. [$F_0$ (day x) – $F_0$ (day 0)] / $F_0$ (day 0), where x is the target day, is used as a measure of cyanobacterial culture growth (Perin et al., 2015). **Figure 6 (a)** shows the $F_0$ incremental ratios for selected days under the solar, K dwarf and dark regimes, which increase for the illuminated samples. The cyanobacterium under K dwarf conditions exhibits a significantly higher $F_0$ incremental ratio with respect to solar light (p < 0.05; **Table S1**). Under continuous darkness, the cyanobacterium exhibits a decrease in $F_0$ incremental ratios towards marginally negative values with time, as the cell concentration at the start of the experiment was higher than towards the end. **Figure 6 (b)** shows the average integrated density (IntD) of the cyanobacterium which is an indicator of culture growth. The results show that the IntD increases with time for the illuminated samples, exhibiting overall higher, but not significant, values in the K dwarf compared to the solar radiation regime (p > 0.05). Cyanobacteria under constant dark conditions failed to exhibit significantly measurable IntD.

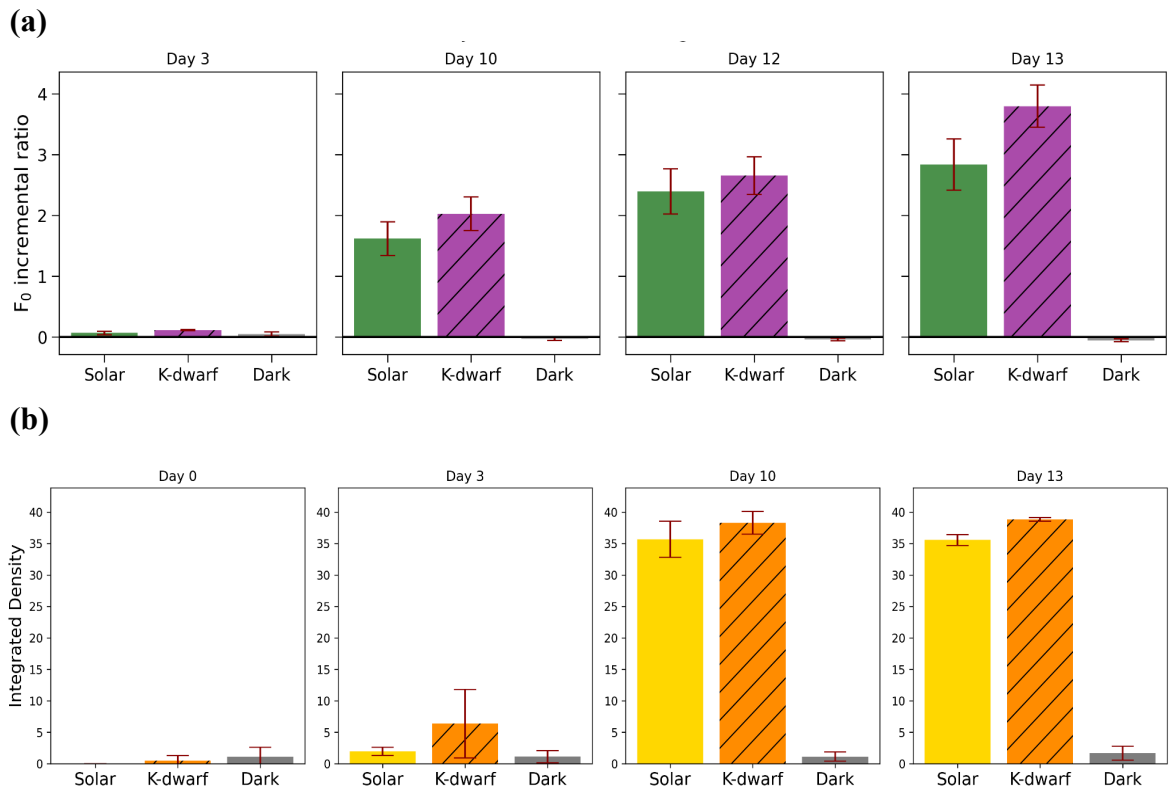

**Figure 6:** $F_0$ incremental ratios and integrated densities of the cyanobacterium on selected days under solar, K dwarf and dark conditions. (a) $F_0$ incremental ratios ([$F_0$ *(day x)* – $F_0$ *(day 0)*] / $F_0$ *(day 0)*), where $F_0$ is the minimum fluorescence value, are proportional to the increment of chlorophyll *a*, and thus, to the number of cells in the considered spot. As such it is used as an indicator of culture growth. The error bar is 1σ. (b) Integrated densities, which were calculated in *Python* by summing up the raw values of all the pixels (RawIntD) within a region of interest (ROI) of an image (see e.g., Fig. 5) and multiplying it by the ratio of the image area in scaled units to the area in pixels: IntD = RawIntD × (Area in scaled units) / (Area in pixels). As such, they are a measure of bacterial culture growth. The error bar is 1σ.

**Table 1.** Photosynthetic efficiency values of garden cress (*Lepidium sativum*) and the cyanobacterium *Chroococcidiopsis* sp. CCMEE 029 under solar and K dwarf radiation measured by a Pulse-Amplitude-Modulation fluorometer (mini-PAM). Fluorescence parameters are as in **Fig. 4**. Fluorescence values for the cyanobacterium experiment are presented only from the seventh day onward, as prior to this time the culture spots are too diluted for reliable measurements. The standard deviation is 1σ. For the K dwarf cyanobacterium sample, three out of four regions of interest (ROI; see **Fig. 5**) showed saturating fluorescence values from the tenth day of the experiment onward (see bold). As a result, these fluorescence values represent the lower limit of measurements and do not include a standard deviation.

| Organism | Time [days] | K dwarf [$F_v/F_m \pm 1\sigma$] | Solar [$F_v/F_m \pm 1\sigma$] |
|---|---|---|---|
| Garden cress (*Lepidium sativum*) | … | … | … |
|  | 3 | 0.739 ± 0.020 | 0.743 ± 0.004 |
|  | 4 | 0.797 ± 0.010 | 0.788 ± 0.021 |
|  | 5 | 0.801 ± 0.007 | 0.791 ± 0.011 |
|  | 6 | 0.797 ± 0.009 | 0.805 ± 0.008 |
|  | 7 | 0.786 ± 0.010 | 0.778 ± 0.007 |
|  | 8 | 0.780 ± 0.022 | 0.778 ± 0.008 |
|  | 9 | 0.781 ± 0.017 | 0.777 ± 0.006 |
| Cyanobacterium (*Chroococcidiopsis* sp. CCMEE 029) | … | … | … |
|  | 7 | 0.311 ± 0.007 | 0.190 ± 0.008 |
|  | 10 | **0.428** ± 0.0 | 0.313 ± 0.016 |
|  | 12 | **0.452** ± 0.0 | 0.402 ± 0.019 |
|  | 13 | **0.459** ± 0.0 | 0.413 ± 0.013 |

## Discussion

Understanding the effects of K dwarf radiation on photosynthesis and growth is of foremost importance not only for the assessment of its viability for phototrophic organisms but also for the interpretation of atmospheric biosignatures outside of the Solar System. However, there has been a dearth of experimental data on the responses of oxygenic photosynthetic organisms to simulated K dwarf light spectra. Most of the focus so far has been on cyanobacteria and plants exposed to the simulated radiation of the cooler counterparts, M dwarf stars (Battistuzzi et al., 2023a, 2023b; Claudi et al., 2020). In this study, to our knowledge, **we present the first experimental data on the responses of photosynthetic organisms to a simulated K dwarf spectrum.**

It is important to keep in mind that the radiation intensities used for the garden cress experiments during the 12-hour daytime period may not fully replicate the natural fluctuations in sunlight intensity experienced by plants in their native environment. Sunlight intensity on Earth varies throughout the day, with peak intensities occurring during the central hours. This variation is crucial for plants to adapt and respond to changing light conditions, including the activation of non-photochemical quenching (NPQ) to mitigate the effects of excess light. Our experimental design prioritizes controlled and reproducible conditions to investigate specific physiological responses under standardized environmental parameters, rather than the ideal growth conditions for a given organism. While our approach may not perfectly mirror real-world conditions, it allows us to isolate and study the effects of the two radiation regimes on plant physiology in a controlled setting. Previous studies have indicated that both shortwave and longwave radiation are necessary for the ideal growth of garden cress, which absorbs primarily in both the red (around 665-680 nanometers) and blue (around 430-450 nanometers) regions of the electromagnetic spectrum (Ajdanian et al., 2019; Guidi et al., 2017; Ménard et al., 2006; Okamoto et al., 1996). Our results, which demonstrate slightly higher dry masses for garden cress under solar light, align with the biochemical parameters of the plant. This alignment is further supported by the fact that solar light exhibits higher fluxes in the shortwave regime compared to K dwarf stars (see **Fig. 7** in the Methods section). Furthermore, while we acknowledge the relationship between reduced $F_v/F_m$ values and plant stress, it is worth highlighting that the differences observed between garden cress grown under solar and K dwarf conditions are minor and not statistically significant (see **Table 1** and **Table S1** in the Supplementary Materials). Consequently, despite the association, these slight variances in $F_v/F_m$ values do not conclusively indicate a substantial level of stress experienced by the garden cress under solar radiation compared to K dwarf conditions.

Temperature measurements conducted throughout the experiments show stable conditions within the temperature ranges conducive to the growth of garden cress (Sattari Vayghan et al., 2020) (see **Fig. S3** in the Supplementary Materials). Despite small average temperature variations observed between experiments under the three radiation regimes, these variations do not substantially contribute to the differences in plant responses observed across the samples. Despite the differences in spectra, the plant showed the ability to grow similarly under K dwarf radiation compared to the solar regime. This can be attributed to the adequate shortwave and increased longwave fluxes provided by the K dwarf spectrum. Notably, it exhibited this ability despite receiving three times lower photosynthetic photon flux densities (PPFD), which impact biomass accumulation and photosynthetic responses (see **Fig. S1** and **Table 2**) (Deo et al., 2022; Ke et al., 2023, 2021; Welander and Ottosson, 1997). If garden cress, which has evolved under the light of our Sun, can demonstrate comparable growth when exposed to K dwarf radiation, it raises the possibility that garden cress evolved on an exoplanet orbiting a K dwarf star could potentially outperform Earth's garden cress in terms of biomass production. The cyanobacterium examined in this study has developed adaptations to endure various environmental stressors, including high radiation levels, and is commonly found in habitats with limited light availability, such as desert soils or endolithic environments (Billi et al., 2021, 2019, 2017; Casero et al., 2020; Friedmann, 1980). In such conditions, shorter-wavelength light is diminished due to scattering and absorption by the surrounding environment. Consequently, the cyanobacterium may have evolved to effectively utilize longer-wavelength light, which is less impacted by scattering and can penetrate deeper into microbial mats or substrates. Hence, considering that irradiation levels strongly fluctuate in desert environments, it is plausible that the light intensity used in this study for the K dwarf scenario falls within the range of natural irradiance variations and is thus preferred by the cyanobacterium.

These results can bring us closer to addressing which stellar environments could be the optimal candidates in the search for habitable worlds. We must keep in mind, though, that this study is constrained by the number and choice of organisms as well as by the evolutionary differences between them. For example, cyanobacteria evolved some 3.5 billion years ago, a time during which Earth received 0.75 times the solar constant and the atmosphere contained orders of magnitude more carbon dioxide and less oxygen and ozone, causing the surface of the Earth to receive higher amounts of X-ray and UV radiation (Cnossen et al., 2007; Hessler, n.d.; Kaltenegger et al., 2020; Kasting, 1993; Lehmer et al., 2020; Pavlov et al., 2000; Payne et al., 2020; Schopf and Packer, 1987; Walker, 1985). This could have influenced the evolutionary trajectory of cyanobacteria, leading them to adapt to and favor lower-intensity and less energetic light. The K dwarf spectrum, with its reduced levels of radiation, may align better with the light conditions to which cyanobacteria were exposed during their evolutionary history. Garden cress, on the other hand, evolved much more recently, around the time when plants started transitioning from the oceans onto land some 400–300 million years ago (Dunn, 2013). This time was accompanied by a comparable solar constant as well as oxygen, ozone and carbon dioxide levels to today (Dunn, 2013; Glasspool and Scott, 2010; Mills et al., 2021). Garden cress, as a land plant, is influenced by a range of environmental cues that are intricately linked to solar light and has evolved to thrive in such a terrestrial environment where it is exposed to more direct solar light than the cyanobacterium. This can be seen in their photosynthetic pigments which evolved to absorb more shortwave light and is also consistent with the dry mass results under solar irradiance of our study.

In addition to these experiments, the selection of a K dwarf as a host star is among several proposed factors that could enhance the potential habitability of an exoplanet (Arney, 2019; Heller and Armstrong, 2014; Schulze-Makuch et al., 2020). Together with the right obliquity and eccentricity of a planet, these stars would provide the 'goldilocks' stellar environment for the emergence of superhabitable worlds and potential extraterrestrial life (Arney, 2019; Cuntz and Guinan, 2016; Jernigan et al., 2023; Wolfe et al., 2010). Hence, the results of this work are suggestive of the viability of photosynthesis that has evolved with a different parent star and can be used as a stepping stone to simulate signatures of oxygenic photosynthesis on exoplanets around K dwarf stars (Kiang et al., 2007a, 2007b). This can aid in the interpretation of future exoplanetary atmospheric signatures obtained by e.g. JWST or studied by the KOBE or Terra Hunting experiments (Lillo-Box et al., 2022; Naylor, n.d.).

## Conclusion

Garden cress exposed to K dwarf radiation shows comparable responses to those under solar conditions, demonstrating the ability of the plant to grow under this alternative type of radiation. The cyanobacterium demonstrated a significantly more positive response to K dwarf radiation, exhibiting both higher photosynthetic activities and culture growth.

These findings not only highlight the coping mechanisms of photosynthetic organisms to modified radiation environments, but they imply the principal habitability of exoplanets orbiting K dwarf stars. Understanding the responses of organisms to different spectra is not only crucial for assessing the suitability of K dwarfs as host stars for habitable exoplanets, but also for interpreting atmospheric biosignatures beyond our Solar System. To obtain a holistic understanding of the responses to K dwarf radiation, further investigations should focus on the choice of representative organisms from Earth's major biomes, such as aquatic ecosystems, grasslands, forests, deserts, and tundra. Beyond the choice of biological organisms, these experiments could further be enhanced by incorporating an environmental simulation chamber which would allow modification and monitoring of ambient gases, pressures, and temperatures. Additionally, variations in day length as well as a simulated twilight period could be introduced to make the experiments more realistic. The results of these future experiments could make this study a benchmark on a long road of laboratory tests that can be used to prioritize extrasolar planets for costly follow-up observations like transit spectroscopy.

## Methods

**General experimental course**

In this work we conducted experiments to evaluate the viability of K dwarf radiation on photosynthetic organisms using a starlight simulator. The subsequent outline constituted the entirety of the experiments:

   I.   To simulate the stellar spectra, we used an LED stellar simulator with a variable spectrum module which allows for the reproduction of a variety of spectral environments in the 350-1500 nm wavelength range.
   II.  We modeled the emission spectrum of a ~4300 Kelvin K dwarf star (i.e. spectral type: K4.5) using the PHOENIX spectral library (Husser et al., 2013). We calculated the daily globally averaged electromagnetic spectrum of the K dwarf star transmitted to the surface of a hypothetical planet with an Earth-like atmosphere in the center of its habitable zone (Kopparapu et al., 2013). We repeated the calculations for the Sun to obtain the control radiation. Both spectra were used as an input for the LED stellar simulator.
   III. Given the rationale that Earth-like planets in the centers of the habitable zones of most K dwarf stars (i.e. with masses higher than ~0.5 solar masses) lie outside of the radius at which a planet in a circular orbit would be tidally locked (Kasting et al., 1993), we introduce a 12-hour day/night cycle in all of the experiments in the form of a square wave.
   IV.  We built a small external light isolating chamber in which the selected photosynthetic organisms were hosted during the experiments. Ambient conditions were logged with a temperature and humidity sensor.
   V.   Samples of the selected photosynthetic organisms were subsequently exposed to solar, K dwarf as well as continuously dark conditions for 9 and 13 days (for garden cress and cyanobacterium experiments, respectively) and their responses were determined in post-experimental analysis.

**Model spectra**
*Calculating the modeled K dwarf spectrum at the planetary surface*

We calculated the wavelength resolved flux received at the surface of the K dwarf planet by multiplying the wavelength resolved top-of-the-atmosphere (TOA) flux with the transmission (telluric) spectrum of Earth provided to us by Baker et al.(Baker et al., 2020). Their code wraps the *Planetary Spectrum Generator* (PSG) radiative transfer tool in order to calculate the Earth spectra and has a lower limit at 200 nm (Villanueva et al., 2018). This transmission spectrum was calculated at zenith over Göttingen, Germany on a random fall day and reveals the composition of Earth's atmosphere, with its prominent water vapor absorption bands (see **Fig. 7**). The transmission spectrum of Earth serves as a first proxy towards assessing surface conditions of potentially habitable exoplanets. **To our knowledge, we calculated for the first time the stellar spectrum of a K dwarf star transmitted to the surface of a hypothetical habitable zone planet with an Earth-like atmosphere using Earth's telluric spectrum** (see upper panel of **Fig. 7**). Analogously, we modeled the emission spectrum of the Sun using the PHOENIX spectral library and placed the Earth at 1 AU, where it receives one solar constant $S_0$. We calculated the emission spectrum at TOA as well as the solar daily averaged spectrum transmitted to the surface of Earth using Earth's telluric spectrum (see lower panel of **Fig. 7**). A numerical summary for the Earth–Sun and Exoplanet–K dwarf systems is presented in **Table 2**.

**Creating the starlight simulator**

For the starlight simulator we used the Pico LED small area solar simulator with the *Variable Spectra Module* (VSM) from the manufacturer *G2V Optics Inc*. The VSM allows for software-controlled intensity manipulation of up to 30 LED channels, which are treated as linear superpositions of up to three skew gaussian functions, in the wavelength range 350 – 1500 nm, to replicate a variety of solar as well as other stellar conditions.

We produced a nonlinear fit to the modeled PHOENIX surface spectra which we used as input to communicate with the individual LED channels of the *Pico* software (see 'simulated' spectra in **Fig. 8**). We used *Python* to visualize the produced fits in comparison to our modeled spectra and manually further tweaked the fits to minimize the discrepancies. The final *Pico* fits (i.e., 'simulated spectra') and the corresponding modeled surface spectra are shown in **Fig. 8**. An automatic day/night cycle (i.e., 12-hour cyclical switching on and off the lamp) was also implemented in *Python* in the form of a square wave to ensure a more consistent experimental progression.

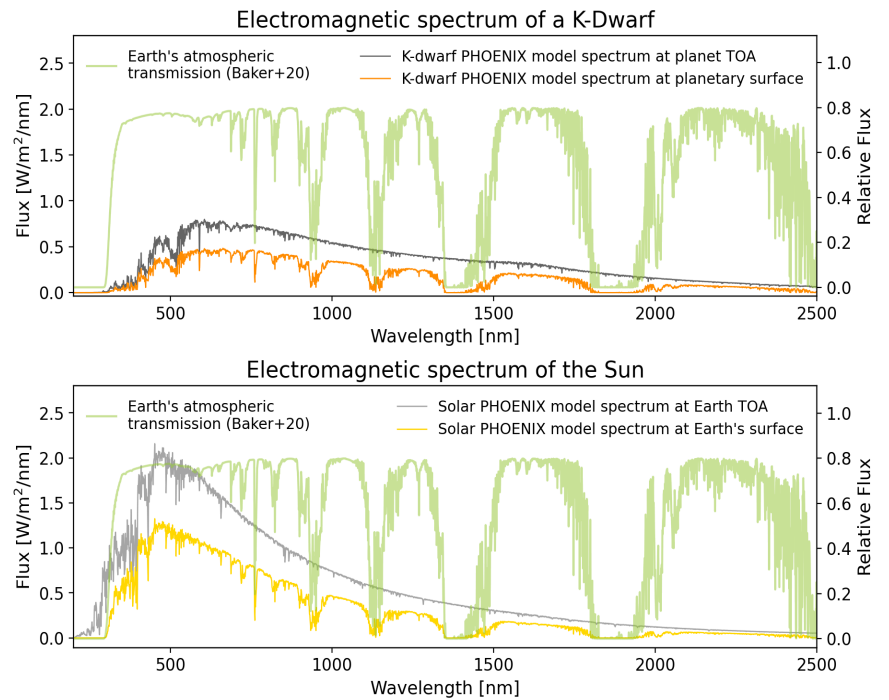

**Figure 7:** High-resolution spectra of the K dwarf star (upper panel) and the Sun (lower panel). The green spectrum in both panels is Earth's telluric (transmission) spectrum provided by Baker et al. (Baker et al., 2020). Upper panel: the dark gray spectrum is the wavelength resolved top of the atmosphere (TOA) flux at ~0.441 AU from the host star calculated in this work. The orange spectrum is the daily averaged model spectrum of a K dwarf star transmitted to the surface of our hypothetical planet with an Earth's telluric spectrum. The daily averaging was calculated by assuming a 12-hour daylight cycle (see Methods). Lower panel: Same as in upper panel, but for the Earth/Sun system.

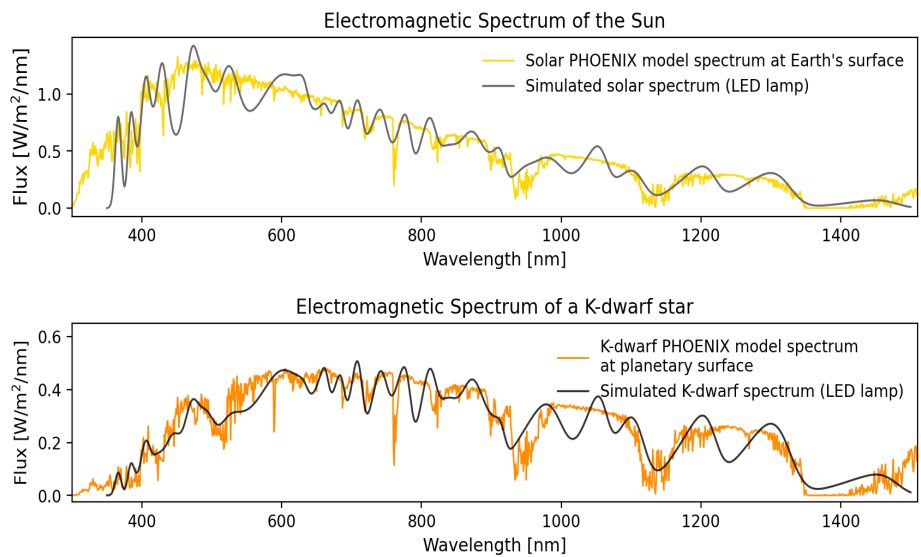

**Figure 8:** Modeled surface spectra from Figure 7 and their corresponding LED lamp simulations. The lamp has thirty tunable LED channels which are treated as linear superpositions of up to three skew gaussian functions whose intensities can be manipulated individually. The simulated spectra are used as input to communicate with the LED software.

**Table 2.** Summary of stellar and planetary parameters used in this work for the Sun–Earth and K dwarf–exoplanetary systems. The hypothetical exoplanet lies at 0.441 AU from its host K dwarf star, placing it in the center of the habitable zone whose inner and outer boundary are given by the moist and maximum greenhouses, respectively. For a derivation of the values, see the Methods section and **Supplementary Materials**. $S_0$ stands for the solar constant.

|  | **Earth/Sun** | **Exoplanet/K-star** |
|---|---|---|
| Stellar type | G2V | K4.5 |
| Stellar effective temperature | 5772 K | 4285 K |
| Stellar Mass | 1 $M_\odot$ | 0.667 $M_\odot$ |
| Stellar Luminosity | 1 $L_\odot$ | 0.116 $L_\odot$ |
| Stellar Radius | 1 $R_\odot$ | 0.620 $R_\odot$ |
| Star - planet distance | 1 AU | 0.441 AU |
| Effective flux at planetary top-of-atmosphere | 1368 W/m² (= 1 · $S_0$) | 820.8 W/m² (= 0.6 · $S_0$) |
| Effective flux at planetary surface | 1115.4 W/m² | 621.2 W/m² |
| Photosynthetic Photon Flux Density (PPFD) for Garden cress | 1446 µmol m$^{-2}$s$^{-1}$ | 512 µmol m$^{-2}$s$^{-1}$ |
| Photosynthetic Photon Flux Density (PPFD) for Cyanobacterium | 30 µmol m$^{-2}$s$^{-1}$ | 10 µmol m$^{-2}$s$^{-1}$ |
| Red to Far-Red photon flux ratio (R:FR) | 1.28 | 1.16 |

## Laboratory procedures

### Garden Cress: Dry matter accumulation and water intake

Garden cress seeds (*Lepidium sativum*) were exposed to simulated solar and K dwarf radiation and the dry and fresh masses of the seedlings were measured to determine the changes in biomass through time. The garden cress is a rapidly growing annual herb which is adaptable to diverse climates and soil conditions, and as such makes an excellent astrobiological target (Nehdi et al., 2012; Wadhwa et al., 2012). The experiment was also run under dark conditions to ensure that the relative growth of garden cress under K dwarf and solar conditions is in fact due to radiation and not due to the intrinsic availability of nutrients in the plant seeds themselves. The irradiated samples were positioned at a stand at 3 cm to the base of the setup, equaling 8 cm to the irradiation source. Sand was used as the substrate as it allowed for the complete daily extraction of the sample, i.e., of the sprouted seedlings and the entirety of their roots. The methodology of measuring the fresh and dry mass of the sample is based on several works for measuring biomass accumulation ("Dry mass change during germination of bean seeds," n.d.; Filson, 2005; Hoh, 2002). Four independent biological replicates of garden cress were used for the experiments.

The following steps were implemented:
   **1) Control group:**
   A sample of **ninety control garden cress seeds** was used.
   a) The fresh masses of 9 groups à la 10 control seeds were measured on a digital $1.0 \times 10^{-4}$ g precision scale. The mean and standard deviation for the nine groups of ten control seeds was calculated.
   b) The nine groups à la ten seeds were desiccated by drying in an oven at 40℃ for two to three days and subsequently in a vacuum desiccator until a constant dry mass was reached. This was the case after a further two to three days.
   c) The final dry masses of nine groups à la ten control seeds were measured, and the mean and standard deviation were calculated once more.

   **2) Experimental group:**
   A further sample of **ninety experimental garden cress seeds** was used. The following steps were performed for the solar and K dwarf radiation sources implementing a 12-hour day/night cycle, as well as for dark conditions.
   *Pre-experiment*
   a) The fresh masses of 9 groups à la 10 experimental group garden cress seeds were measured on the digital $1.0 \times 10^{-4}$ g precision scale. The mean and standard deviation for the nine groups of ten control seeds was calculated.
   b) Those ninety seeds were soaked and rinsed with tap water.
   c) A sand substrate was placed in a 5cm-diameter Petri dish and wetted with water.
   d) The ninety rinsed seeds were initially placed on the wet sand using laboratory tweezers and then placed under the radiation (or in a closed 40 x 40 x 20 cm box for the dark scenario).
   *During experiment*
   e) Temperatures and relative humidity levels were tracked and logged with a temperature- and humidity sensor throughout the full duration of the experiment (see **Fig. S3** in the Supplementary Materials).
   f) The plants were watered daily with a constant amount of water, ensuring consistent hydration levels throughout the duration of the experiments.
   g) A random selection of ten seeds, including roots, was extracted from the population every 24 hours for a total of 9 days – the interval between consecutive measures remained equal for consistency.
   h) The extracted seeds/seedlings/plant matter were thoroughly rinsed to remove sand remnants.
   *Post-experiment*
   i) The fresh masses of the extracted material were measured on the digital $1.0 \times 10^{-4}$ g precision scale every 24 hours.
   j) The extracted material was desiccated until a constant mass was reached, including 2–3 days in the oven at 40℃ and a further 2–3 days in the vacuum desiccator.
   k) The final dry masses of the extracted seeds/seedlings/plant matter were measured.

Although the experiments conducted under different radiation regimes share the confounding variable of seed proximity, potentially resulting in shading and altered growth patterns due to the rapid germination of the first seeds (Casal, 2013), the random selection of seedlings for analysis in both scenarios facilitates a comparative evaluation of the plant's reaction to the modified light conditions. While acknowledging the potential influence of seed proximity on plant growth, the random sampling approach helps mitigate any systematic biases associated with this shared variable. Consequently, we contend that the observed differences in plant responses predominantly reflect the effects of the varying light conditions.

**Garden Cress: Photosynthetic Efficiency**
The photosynthetic efficiency of garden cress under solar-, K dwarf and dark conditions was measured with the Mini-PAM II (Pulse-Amplitude-Modulation, Walz, Germany) fluorometer which determined chlorophyll *a* (Chl *a*) fluorescence yields. Chl *a* fluorescence yields serve as an indicator for photosynthetic performance and are often used to identify chemical or environmental stressors (Ogren, 1990; Percival, 2005; Percival and Sheriffs, 2002). In this work, the fiber optics tip of the Mini-PAM was placed at 2 cm to sample level and was aligned at a 60° angle relative to the measuring plane with a 1 cm-diameter measuring area. The maximum photochemical quantum yield ($F_v/F_m$) of the photosystem II (PSII) is given by the equation: $F_v/F_m = (F_m - F_0) / F_m$, where $F_m$ is the maximum fluorescence level in mV generated by a 0.6 second saturation pulse (SP) which closes all PSII reaction centers and $F_0$ is the minimum fluorescence level in mV excited by a very low-intensity measuring light (ML; frequency = 5 Hz) to keep PSII reaction centers open (Genty et al., 1989; Kitajima and Butler, 1975). A ML intensity setting of 12 corresponds to 6000 µmol m$^{-2}$s$^{-1}$ at a 7 mm fiber-optics-to-sample-level distance and is adjusted at increments of 500 µmol m$^{-2}$ s$^{-1}$. The final PAM settings can be found in **Fig. S2**. $F_v/F_m$ was measured on a sample which had been acclimated to the dark for 25-30 minutes.

To ensure a denser plant population and a more uniform surface area, we conducted the photosynthetic efficiency experiments by sprinkling a handful of garden cress seeds onto a sterile cotton substrate. We opted for sterile cotton as the substrate because the weight and adhesive properties of wetted sand grains would have hindered the growth of the cress seeds. All measurements under the PAM settings were obtained from four independent biological replicates of garden cress under solar, K dwarf and dark conditions. The parameter values were measured every 24 hours for 9 days per illumination-type and biological replica.

**Cyanobacteria: Photosynthetic Efficiency and Integrated Density**

We used the desert cyanobacterium *Chroococcidiopsis* sp. CCMEE 029 (Culture Collection of Microorganisms from Extreme Environments) extracted from the Negev Desert in Israel for this portion of the study. Cyanobacteria are believed to contribute to 20-30% of the Earth's primary photosynthetic productivity and have a collective worldwide biomass of approximately $3 \times 10^{14}$ grams of carbon (Garcia-Pichel et al., 2003; Pisciotta et al., 2010). As such, this strain makes an excellent astrobiological target which can withstand a variety of extreme conditions and can be found in numerous environments (Billi et al., 2019, 2017; Cumbers and Rothschild, 2014). We grew the cyanobacterium culture in a liquid BG-11 medium with a 10$^{-1}$ dilution in a 25°C incubator without shaking under a 4000 Kelvin cold LED lamp until it reached an exponential growth phase. At this point we transferred the culture to a BG-11 Agar plate in 20 microliter-sized drops. Given that this cyanobacterial strain exhibits an optimal growth at intensities much smaller than garden cress, i.e. PAR intensities of 30–40 µmol (photons) m$^{-2}$s$^{-1}$, we decreased the wavelength-distributed flux of the modeled solar emission spectrum as well as its corresponding simulated LED spectrum by 48 times. To keep the original photonal ratio of the two spectral types, the K dwarf PAR intensity was decreased to 10 µmol (photons) m$^{-2}$s$^{-1}$ (see **Table 2**). We then exposed the Agar plate to these radiation regimes for 13 days, implementing a 12-hour day/night cycle. Every few days we documented the visual changes of the culture with a camera. Four independent biological replicates of the cyanobacterial strain were used for the experiments.

We measured the photosynthetic efficiencies of dark-acclimated samples with the mini-PAM fluorometer. We used Gain level 3 and a ML intensity of 12 in the PAM settings to capture the low photosynthetic signal of the cyanobacterium spots in the initial phases of the experiments. The remaining settings were as for the garden cress experiments (see **Fig. S2**). We calculated the corresponding integrated densities (IntD) with *Python* by summing up the raw values of all the pixels (RawIntD) within a region of interest (ROI) in the image (see e.g., **Figure 5** in the main text) and multiplying it by the ratio of the image area in scaled units to the area in pixels:

$$IntD = RawIntD * (Area\ in\ scaled\ units) / (Area\ in\ pixels) \qquad (1)$$

We used an RGB (red, green, blue) pixel range of ±10 around the average RGB values of all regions of interest as a color threshold and calculated the integrated densities of each spot. The threshold captures all tonalities of pixels in the samples in accordance with their contribution to the culture growth.

**Acknowledgements**
The first author would like to thank the scholarship organization *Studienstiftung des deutschen Volkes* without which this work would not have been possible. The authors also thank professor Dr. Bernd Szyszka and his Ph.D. student Sri Hari Bharath Vinoth Kumar at the Technische Universität Berlin for providing the essential *Pico* LED small area solar simulator from the manufacturer *G2V Optics Inc.* for the experiments. The authors also thank Dr. Ashley D. Baker for providing us with the invaluable telluric spectrum of Earth. Furthermore, the authors thank Dr. Nicoletta La Rocca and her research group at the University of Padova for invaluable input and advice concerning the experimental setup of the cyanobacterial portion of this work.
   **Funding**
   *Studienstiftung des deutschen Volkes* doctoral scholarship (IV)
   PLATO Data Center grant 50OO1501 (RH)
   **Author contributions**
   Conceptualization: DSM
   Methodology: IV, DSM, RH
   Investigation: IV
   Visualization: IV
   Formal analysis: IV, RH
   Funding acquisition: IV
   Supervision: DSM, RH
   Writing – original draft: IV
   **Data and materials availability**
   All data generated or analyzed during this study are included in this published article and their sources are cited where applicable.
   **Competing interests**
   The authors declare no competing interests.
   **Correspondence and requests for materials should be addressed to**
   Iva Vilović
**Supplementary Materials**
Yes